\documentclass[aps, prd, notitlepage, letterpaper, 11pt, amsmath, amssymb, amsfonts, floatfix, nofootinbib, superscriptaddress]{revtex4-1}

\usepackage[utf8]{inputenc}
\usepackage[english]{babel}
\usepackage{microtype}
\usepackage{graphicx} 
\usepackage{dcolumn}
\usepackage{slashed}
\usepackage{lineno}
\usepackage{color}
\usepackage{amsmath}
\usepackage{amssymb}
\usepackage[dvipsnames]{xcolor}
\usepackage{cancel}
\usepackage{framed}
\usepackage{comment}
\usepackage{mathtools}
\usepackage{hyperref}
\usepackage{indentfirst}
\usepackage{multirow}
\usepackage{physics}

\usepackage{ulem}

\definecolor{added}{rgb}{0,0,1}
\definecolor{deleted}{rgb}{1,0,0}

\newcommand{\overbar}[1]{\mkern 1.5mu\overline{\mkern-1.5mu#1\mkern-1.5mu}\mkern 1.5mu}

\makeatletter
\newcommand*{\toccontents}{\@starttoc{toc}}
\makeatother

\allowdisplaybreaks

\begin{document}

\title{Doubly Blind Spots in Scalar Dark Matter Models}

\def\ucsc{Department of Physics and Santa Cruz Institute for Particle Physics, University of California, Santa Cruz, CA 95064, USA}

\author{Wolfgang~Altmannshofer} \email[]{waltmann@ucsc.edu}
\affiliation{\ucsc}

\author{Brian~Maddock} \email[]{bmaddock@ucsc.edu}
\affiliation{\ucsc}

\author{Stefano~Profumo} \email[]{sprofumo@ucsc.edu}
\affiliation{\ucsc}

\begin{abstract}
We consider a framework where the Standard Model is augmented by a second $SU(2)$ scalar doublet and by a real scalar singlet that, protected by a $Z_2$ symmetry, provides a particle Dark Matter candidate. We show that this setup allows for {\it doubly blind spots} at both collider searches for anomalies in the Higgs invisible decay width, and at direct Dark Matter detection. The blind spots originate from cancellations between interfering diagrams featuring different neutral scalar exchanges, and from cancellations driven by the two-Higgs doublet structure in the vertex coupling the singlet state with the Standard-Model-like Higgs. We demonstrate that the blind spots arise in a wide and generic array of realizations for the two-Higgs doublet model, including scenarios with a non-trivial flavor structure. We provide analytical formul\ae ~that describe the location of the blind spots in the theory parameter space, and we discuss the resulting phenomenology.
\end{abstract}

\maketitle

\section{Introduction}

The physical origin and nature of the dark matter (DM) that permeates the universe and underpins the formation and evolution of structure, is at present unknown (see e.g. \cite{Bertone:2010zza, Profumo:2017hqp} for a review). A compelling possibility is that of 
weakly interacting massive particles (WIMPs)\cite{deSwart:2017heh}, new massive particles, neutral or close-to-neutral under electromagnetic and strong interactions, but possibly charged, or mixed with particles which are charged under weak interactions. Perhaps the most minimal possibility of the former is a new, very massive SU(2) multiplet (a possibility known as ``minimal DM'', \cite{Cirelli:2005uq}), and of the latter is a new real scalar singlet ($S$) that interacts via a quartic coupling with the standard model (SM) Higgs \cite{McDonald:1993ex,Profumo:2007wc, Profumo:2010kp, Boucenna:2011hy, Wainwright:2012zn, Profumo:2014opa, Feng:2014vea}. Here, we will be concerned with this second possibility.

Part of the appeal of WIMPs is that they are, generically, in thermal equilibrium in the early universe, eventually decoupling (``{\it freezing out}'') with an abundance that often is in the correct range to explain the observed amount of DM in the universe. While a far-ranging program of direct and indirect searches for WIMPs has been under way for decades now, no conclusive signal has yet been reported \cite{FermiLAT, MAGIC, Holder:2008ux, Aharonian:2006pe,Akerib:2016vxi}: strong constraints exist as a result. In particular, the simple possibility of a singlet scalar mentioned above (SM+S) is very strongly constrained, see e.g. \cite{Arcadi:2017kky}.

Here, we consider a slight extension of the singlet scalar DM model to a framework where the SM is enriched with a second Higgs doublet (two-Higgs doublet model, or 2HDM) in addition to the real scalar singlet (we thus dub this scenario 2HDM+S). This case's phenomenology is significantly richer, ultimately because multiple new particles now couple to the DM candidate. In particular, here we are concerned with the possibility that {\it blind spots} arise in the 2HDM+S as a result of either (i) destructive interference between diagrams involving different neutral scalars, or (ii) exact cancellations in the $S$ coupling to the SM-like Higgs. 

Blind spots have been pointed out in the literature before, see e.g. \cite{Cheung:2012qy, Huang:2014xua, Han:2016qtc, Han:2018gej, Crivellin:2015bva, Badziak:2015exr, He:2008qm, He:2011gc, He:2013suk, He:2016mls, Chang:2017gla, Chang:2017dvm} in the context of supersymmetry, and \cite{Greljo:2013wja, Arcadi:2018pfo} in the context of a two-Higgs doublet model. However, the central point we make here is that the blind spots in the 2HDM+S scenario are largely generic, i.e. they arise in a wide variety of realizations for the implementation of the specific 2HDM; secondly, we point out that blind spots pertain to both collider searches (where one looks for a deviation of the invisible Higgs decay width from standard expectations) and direct DM detection, and that, on occasion, the two blind spots can overlap.

The outline of our study is as follows: In sec.~\ref{model} we lay out the basic ingredients and parameters of the 2HDM+S model that we will discuss. In sec.~\ref{cancellations} we give an in-depth look at how to form blind spots in a generic 2HDM+S setup. In sec.~\ref{pheno} we consider the broader implication of these blind spots and how they can open up parameter space that has previously been ruled out. Finally, we conclude in sec.~\ref{conclusion}.

\section{Generic Two Higgs Doublet Model + Singlet}
\label{model}
\begin{table}
\begin{center}
\begin{tabular}{l cccccc}
\hline\hline
Model & ~~~~~$u, c$~~~~~ & ~~~~~$t$~~~~~ & ~~~~~$d,s$~~~~~ & ~~~~~$b$~~~~~ & ~~~~~$e,\mu$~~~~~ &  ~~~~~$\tau$~~~~~ \\ 
\hline
Type 1A & $\phi_1$& $\phi_1$&$\phi_1$& $\phi_1$& $\phi_1$&$\phi_1$ \\
Type 1B & $\phi_2$& $\phi_1$&$\phi_2$& $\phi_1$& $\phi_2$&$\phi_1$ \\[8pt]
Type 2A& $\phi_1$& $\phi_1$& $\phi_2$& $\phi_2$& $\phi_2$& $\phi_2$\\
Type 2B& $\phi_2$& $\phi_1$& $\phi_1$& $\phi_2$& $\phi_1$& $\phi_2$\\[8pt]
Flipped A& $\phi_1$& $\phi_1$&$\phi_2$& $\phi_2$& $\phi_1$&$\phi_1$ \\
Flipped B& $\phi_2$& $\phi_1$&$\phi_1$& $\phi_2$& $\phi_2$&$\phi_1$ \\[8pt]
Lepton-Specific A &$\phi_1$& $\phi_1$& $\phi_1$ &$\phi_1$& $\phi_2$& $\phi_2$\\
Lepton-Specific B &$\phi_2$& $\phi_1$& $\phi_2$ &$\phi_1$& $\phi_1$& $\phi_2$ \\
\hline\hline
\end{tabular}
\end{center}
\caption{Summary of the way in which the SM quarks and leptons couple to the Higgs doublets $\Phi$ and $\Phi^\prime$  in each of the considered models. In the models with natural flavor conservation~(A), all three generations of each fermion type couple to the same Higgs doublet. In the flavorful models~(B), the first two generations and the third generation couple to different Higgs doublets.}
\label{PhiCouplings}
\end{table}

We consider a 2HDM with a generic flavor structure, augmented with a real scalar singlet charged under a $Z_2$ discrete symmetry: This means that we extend the SM with a second Higgs doublet, that can also couple to the SM fermions, as well as a scalar singlet. We consider the most generic Lagrangian for a 2HDM~\cite{Branco:2011iw}. The part of the Lagrangian that describes the Yukawa couplings of the two Higgs doublets, $\phi_1$ and $\phi_2$, with the SM fermions in a generic 2HDM looks like 
\begin{align}
\label{eq:yuk}
-\mathcal{L}_\text{Yuk} = \sum_{i,j} \bigg( \lambda^u_{ij}&(\overbar{Q}_i u_j)  \tilde{\phi_1} +  \lambda^d_{ij}(\overbar{Q}_i d_j) \phi_1 +  \lambda^e_{ij}(\bar{\ell}_i e_j) \phi_1 \nonumber \\ 
& +\lambda^{\prime u}_{ij}(\overbar{Q}_i u_j) \tilde{\phi_2} +  \lambda^{\prime d}_{ij}(\overbar{Q}_i d_j) \phi_2 +  \lambda^{\prime e}_{ij}(\bar{\ell}_i e_j) \phi_2 \bigg) +\text{ h.c.} ~,
\end{align}
for $\phi_1$ the ``SM-like'' SU(2) doublet and $\phi_2$ the additional doublet, and $\tilde \phi_i = i \sigma_2 \phi_i^*$. After electroweak symmetry breaking, assuming no CP violation, the 2HDM results in five physical Higgs bosons: a light neutral scalar $h$ (which we identify with the 125~GeV Higgs), a heavy neutral scalar $H$ (heavy Higgs), a pseudo-scalar $A$, and two charged Higgs bosons $H^\pm$. 

The flavor structures in 2HDMs are determined by the choice of Yukawa matrices $\lambda^u$, $\lambda^{\prime u}$, and $\lambda^d$, $\lambda^{\prime d}$, which then fix the couplings of the fermions to the various Higgs bosons. Common choices of these Yukawa matrices leads to four well studied models with ``natural flavor conservation'': type~1A, type~2A, flipped~A, and lepton-specific~A. The common aspect of these models is that they avoid flavor-changing neutral currents at tree-level. This is achieved by coupling all three generations of one type of fermion to the same Higgs doublet~\cite{GlashowWeinberg}. The four ways that this can be done are shown in tab.~\ref{PhiCouplings}. 

One can also construct models that allow for flavor-changing neutral currents while being experimentally consistent, such as Flavorful 2HDMs (F2HDM)~\cite{Altmannshofer:2015esa, Altmannshofer:2016zrn, Altmannshofer:2018bch, Altmannshofer:2019ogm} (for related models see e.g. \cite{Das:1995df, Blechman:2010cs, Ghosh:2015gpa, Botella:2016krk}). F2HDMs differ from the models with natural flavor conservation in that they treat the third generation independently from the first two generations; this means that the third generation fermions couple dominantly to the opposite Higgs doublet than their first and second generations counterparts. There are four F2HDMs, mirroring the four flavor diagonal models, referred to as: type~1B, type~2B, flipped~B, and lepton-specific~B. For an in-depth discussion see~\cite{Altmannshofer:2018bch}. The flavor structure of the four flavorful models are summarized in tab.~\ref{PhiCouplings}. 

As we will see, the flavor structure of the quarks has the largest impact on the phenomenology of the DM in these models. Of the eight possibilities discussed above (four ``type A'' models and four ``type B'' models) there are only four different ways to couple the quarks. Both up- and down-type quarks can be coupled in the same way (type~1A/B, lepton-specific~A/B) or they can be coupled in the opposite way (type~2A/B, flipped~A/B). Therefore, without loss of generality, for this analysis we will focus on the type~1A/B and type~2A/B models as they represent the four unique ways to couple the quarks. 

The characteristic pattern of Higgs couplings to the SM quarks in the different types of 2HDMs is determined by two angles: $\alpha$ and $\beta$, where $\alpha$ is the mixing between the two neutral scalar components of the doublets $\phi_1$ and $\phi_2$, and $\tan\beta = v_1/v_2$ is the ratio of the vacuum expectation values of $\phi_1$ and $\phi_2$.
The corresponding terms in the Lagrangian which contain the physical scalar Higgs bosons and quarks can be written as
\begin{equation}
    \mathcal{L} \supset \sum_q ~\bar{q}q~ (y_{q,h}\, h + y_{q,H}\, H),
\end{equation}
where $y_{q,h}$ and $y_{q,H}$ represent the flavor diagonal couplings of the quarks $q$ to the SM-like and heavy Higgs, respectively. 
As discussed, these couplings are characteristic for a given type of 2HDM. Concretely, in our four example scenarios, the couplings of the SM-like Higgs can be expressed in terms of $\alpha$ and $\beta$ as 
\begin{subequations}
\label{SMCouplings}
\begin{eqnarray}
y_{t, h} &=& \frac{m_t}{v} \frac{\cos\alpha}{\sin\beta} ~~~~~~~~~~~~~~~~~~~\text{all types} \\
 y_{b, h} &=& ~~~ \frac{m_b}{v} \times \begin{cases}~~ \frac{\cos\alpha}{\sin\beta} ~~~~~~~~\text{type~1A~,~1B}\\ -\frac{\sin\alpha}{\cos\beta}  ~~~~~~~~\text{type~2A~,~2B} \end{cases}~, \\
 y_{c(u), h} &=& \frac{m_{c(u)}}{v} \times \begin{cases} ~~ \frac{\cos\alpha}{\sin\beta} ~~~~~~~~\text{type~1A~,~2A}\\ -\frac{\sin\alpha}{\cos\beta}  ~~~~~~~~\text{type~1B~,~2B} \end{cases}~, \\
  y_{s(d), h} &=& \frac{m_{s(d)}}{v} \times  \begin{cases} ~~  \frac{\cos\alpha}{\sin\beta} ~~~~~~~~\text{type~1A~,~2B}\\ -\frac{\sin\alpha}{\cos\beta}  ~~~~~~~~\text{type~1B~,~2A} \end{cases}~,
\end{eqnarray}
\end{subequations}
where $v = \sqrt{v_1^2 + v_2^2} \simeq 246$\,GeV is the SM Higgs vev.
For the heavy Higgs boson we have 
\begin{subequations}
\label{HeavyCouplings}
\begin{eqnarray}
y_{t, H} &=& \frac{m_t}{v}  \frac{1}{\tan\beta}\frac{\sin\alpha}{\cos\beta} ~~~~~~~~~~~~~~~~~~\text{all types} \\
 y_{b, H} &=& ~~~ \frac{m_b}{v} \times \begin{cases}~~ \frac{1}{\tan\beta}\frac{\sin\alpha}{\cos\beta} ~~~~~~~~\text{type~1A~,~1B}\\~~ \tan\beta \frac{\cos\alpha}{\sin\beta}  ~~~~~~~~\text{type~2A~,~2B} \end{cases}~, \\
 y_{c(u), H} &=& \frac{m_{c(u)}}{v} \times \begin{cases}~~ \frac{1}{\tan\beta}\frac{\sin\alpha}{\cos\beta} ~~~~~~~~\text{type~1A~,~2A}\\ ~~\tan\beta \frac{\cos\alpha}{\sin\beta}  ~~~~~~~~\text{type~1B~,~2B} \end{cases}~, \\
  y_{s(d), H} &=& \frac{m_{s(d)}}{v} \times \begin{cases}~~ \frac{1}{\tan\beta}\frac{\sin\alpha}{\cos\beta} ~~~~~~~~\text{type~1A~,~2B}\\ ~~\tan\beta \frac{\cos\alpha}{\sin\beta}  ~~~~~~~~\text{type~1B~,~2A} \end{cases}~.
\end{eqnarray}
\end{subequations}
Additional small corrections to the couplings are present in the flavorful models (type~1B and type~2B). They are proportional to small ratios of fermion masses and can be found in~\cite{Altmannshofer:2018bch}. 

The other ingredient in the framework we consider here is a real scalar singlet $S$, which is assumed to be charged under a discrete $Z_2$ symmetry. This scalar singlet only interacts with the SM through the ``Higgs portal'', i.e. through gauge-invariant, renormalizable operators of the type $S^2\phi_i^\dagger \phi_j$. The terms in the scalar potential that contain the singlet $S$ are 
\begin{equation}
{\cal V}_S  = m_S^2 S^2 + \lambda_S S^4 + \lambda_{S_1} |\phi_1|^2 S^2 + \lambda_{S_2} |\phi_2|^2 S^2+(\lambda_{S_{12}}\phi_1^\dagger \phi_2+ \lambda^*_{S_{12}} \phi_2^\dagger\phi_1)S^2~.
\end{equation}
We assume that $m_S^2$ is positive such that $S$ does not obtain a vacuum expectation value and the $Z_2$ symmetry that stabilizes $S$ remains unbroken.
The quartic interactions between the singlet $S$ and the doublets $\phi_1$ and $\phi_2$ are parameterized by the real couplings $\lambda_{S_1}$ and $\lambda_{S_2}$ and the in general complex coupling $\lambda_{S_{12}}$.

In order to obtain the couplings of the DM $S$ to the physical Higgs bosons the Lagrangian must be rotated to the mass basis. Defining the interactions with the mass eigenstates as
\begin{equation}
{\cal L} \supset S^2 \left( h\,g_{SSh} + H\,g_{SSH} + A\,g_{SSA} \right) ~,
\end{equation}
we find 
\begin{subequations} \label{eq:gSSh}
\begin{eqnarray}
 g_{SSh} &=& v ( \lambda_{S_1} \sin\beta\cos{\alpha} - \lambda_{S_2} \cos\beta \sin{\alpha} + \text{Re}(\lambda_{S_{12}}) (\cos\beta\cos\alpha - \sin\beta\sin\alpha) ) ~, \\
 g_{SSH} &=& v ( \lambda_{S_1} \sin\beta\sin{\alpha} + \lambda_{S_2} \cos\beta \cos{\alpha} + \text{Re}(\lambda_{S_{12}}) (\sin\beta\cos\alpha + \cos\beta\sin\alpha) ) ~, \\
 g_{SSA} &=& -v \Im(\lambda_{S_{12}}) ~.
\end{eqnarray}
\end{subequations}

In the following we will assume that the tree-level scalar potential conserves CP, and therefore set $\Im(\lambda_{S_{12}}) = 0$, such that there are no couplings between the dark matter and a single pseudoscalar Higgs. This choice has little impact on our main results. The pseudoscalar interactions lead to spin dependent dark matter scattering, and the corresponding bounds are several orders of magnitude weaker than those from spin independent scattering mediated by the scalars.
     
\section{Experimental Constraints and Blind Spots}
\label{cancellations}

We consider four constraints on the framework under consideration: the thermal relic density of the dark matter candidate (which we enforce to be reflective of the universal dark matter density), spin-independent direct detection, indirect detection via gamma-ray observations, and invisible Higgs decays. Other DM models with extended Higgs sectors can also be constrained by di-Higgs + MET searches, but these constraints are weak in this model (\cite{Blanke:2019hpe}).   In this section we show how the parameters of this model can conspire such to create blind spots in the constraints from direct detection experiments and invisible Higgs decays. 

The relic density refers to the abundance of DM particles left over from freeze out in the early universe versus the inferred abundance of cosmological DM. The latter was measured by PLANCK (utilizing other data sets as well) to be $\Omega h^2 = 0.1198 \pm 0.0015$ \cite{Aghanim:2015xee}. Any viable DM candidate must predict the relic density to be no greater than $\Omega h^2$, barring modification to the universe's expansion history. In our model we consider a standard freeze out scenario where the DM is in thermal equilibrium with the SM in the early universe, which we assume to be radiation dominated. At this time the DM can annihilate into SM particles, but eventually falls out of thermal equilibrium leaving behind some relic abundance. 

DM is abundant in many astrophysical objects and the annihilation of DM into SM particles can generically lead to an excess of gamma rays.
Indirect detection searches for signatures of DM in gamma ray spectra, and sets constraints on DM models in the absence of any significant excess over background \cite{FermiLAT, MAGIC, Holder:2008ux, Aharonian:2006pe}. Notice that the annihilation of DM into SM particles is a relevant process for both determining the relic abundance and understanding indirect detection, so these two processes are correlated, even though the relevant center-of-mass energy for the thermal decoupling process is biased at slightly larger values since the decoupling happens at finite temperature. 

\begin{figure}[tb]
\begin{center} 
\includegraphics[width=0.8\textwidth]{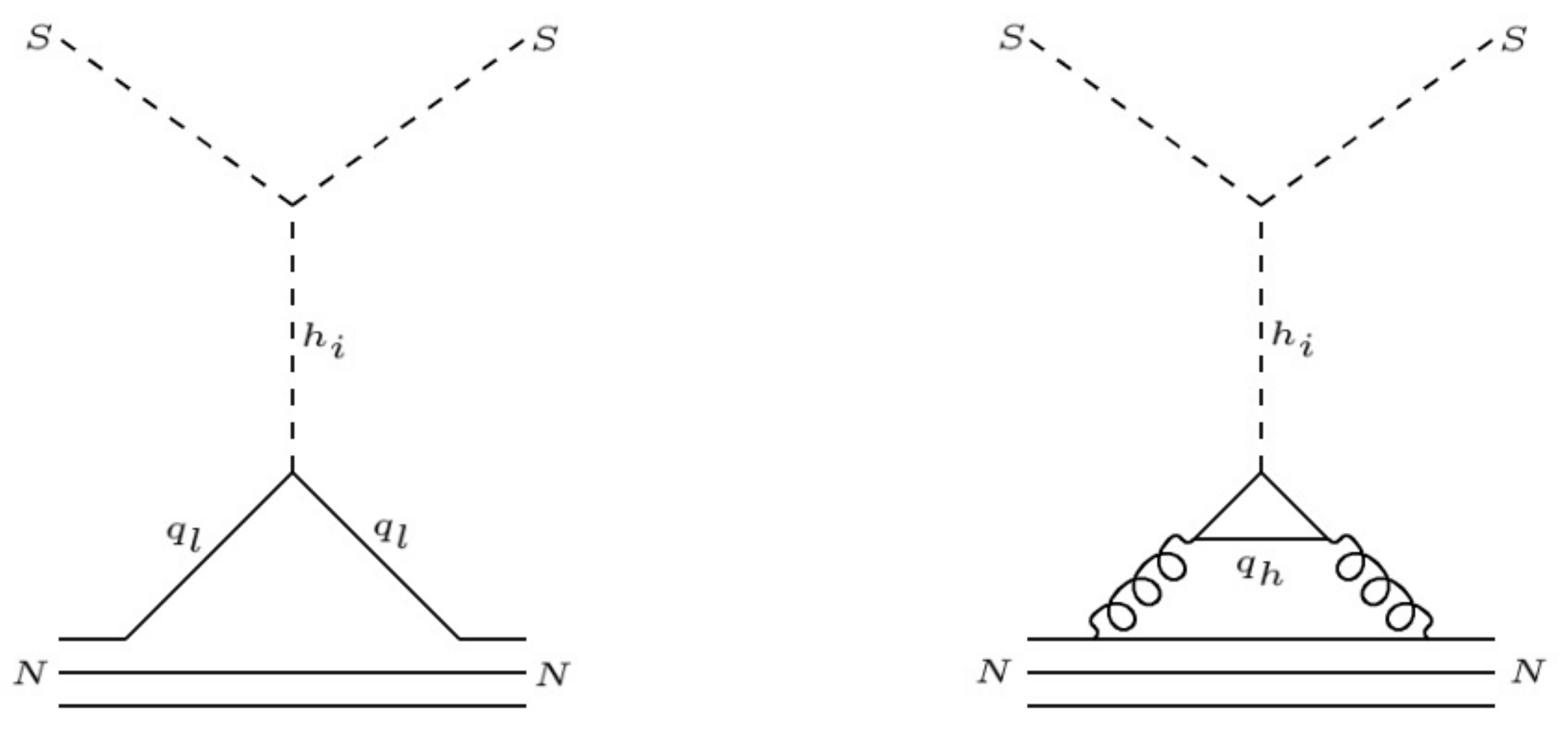}
\caption{The two leading order Feynman diagrams that contribute to the direct detection cross section. Left: tree-level scattering of the singlet $S$, through either the SM-like or heavy Higgs off of light quarks $q_l$. Right scattering of the DM through loops of heavy quarks $q_h$ with the gluons in the nucleon. }
\label{feynDiag}
\end{center}
\end{figure}

Direct detection experiments use nucleons as a target for DM to scatter. When the DM scatters off of nuclei, the latter subsequently recoil; this recoil can then be measured and provides information on the mass and coupling of the DM (see e.g. \cite{Lin:2019uvt} for a recent review). In simple scalar DM models the Higgs mediates the DM-nucleon interaction via direct interaction with the light constituent quarks of the nucleons, or through heavy quark loops with gluons, as shown in fig.~\ref{feynDiag}. The addition of a second Higgs doublet allows for a second mediator to this process and, generically, the scattering amplitudes can destructively interfere, leading to suppression in the direct detection bounds. This is one of the blind spots we consider below. 

Low mass DM, $m_S < \frac{1}{2}m_h$, can also be produced at colliders through the decay of the SM-like Higgs $h \to SS$, which results in an invisible decay of the Higgs. Both the ATLAS and CMS experiments are searching for invisible Higgs decays and are setting bounds on the Higgs to invisible branching ratio~\cite{Sirunyan:2018owy,Aaboud:2019rtt}. The most stringent direct bound comes from CMS and reads BR$(h \to \text{invisible}) < 19\%$ \cite{Sirunyan:2018owy}. The decay rate of $h \to SS$ is determined in large part by the coupling of $g_{SSh}$, the effective coupling of the Higgs to the DM. Similarly to direct detection, there exist regions of parameter space in our model which make $g_{SSh}$ small, effectively avoiding invisible Higgs constraints. This is the second type of blind spot we will consider. 

\subsection{Blind Spots in Direct Detection}

First, we consider the blind spot in direct detection experiments. The spin-independent DM scattering cross section ($\sigma_\text{DM}^\text{SI}$) on a nucleon $N$ reads
\begin{equation}
\sigma_\text{DM}^\text{SI} =\frac{1}{8\pi (m_N + m_S)^2} \Bigg |\sum_{X=h,H} \frac{g_{SSX} m_N^2}{m_{X}^2}   \Bigg(\sum_{q = u,d,s} y_{q,X} f_{Tq} + \sum_{q = c,b,t} \frac{2}{27} y_{q,X}f_{TG} \Bigg) \Bigg |^2.
\label{eq:directdetection}
\end{equation} 
The parameters $y_{q,X}$ represent the couplings of the quarks to the SM-like Higgs and heavy Higgs and are given in eqs.~(\ref{SMCouplings}) and~(\ref{HeavyCouplings}). The couplings of the DM to the Higgs bosons, $g_{SSX}$, are given in eq.~(\ref{eq:gSSh}). The parameters $f_{Tq}$ and $f_{TG}$ represent the nucleon form factors for the quarks interacting with the nucleons in the detector \cite{Belanger:2008sj}, other calculations of these parameters can be found in \cite{Hoferichter:2015dsa, Hoferichter:2017olk}. Blind spots occur for $\sigma_\text{DM}^\text{SI} = 0$, so we must have that
\begin{equation}
\frac{g_{SSh}}{g_{SSH}} \frac{m_H^2}{m_h^2} = -\frac{\sum_{q = u,d,s} y_{q,H} f_{Tq} + \sum_{q = c,b,t} \frac{2}{27} y_{q,H}f_{TG} }{\sum_{q = u,d,s} y_{q,h} f_{Tq} + \sum_{q = c,b,t} \frac{2}{27} y_{q,h}f_{TG}}~.
\end{equation}

Note that to obtain this condition no statement has been made about the flavor structure of the 2HDM. Therefore, this cancellation is a generic feature of 2HDM+S models and is ultimately fixed by the choice of quartic scalar couplings $\lambda_{S_1}$, $\lambda_{S_2}$, and $\lambda_{S_{12}}$, the flavor structure (``type'') of 2HDM, and the 2HDM parameters $\alpha$, $\beta$, and $m_H$. Although this is a generic feature of any flavor structure, here we focus on the type~1A, type~1B, type~2A, and type~2B structures. As discussed above, the type~1A/B and type~2A/B models represent the four ways of coupling the quarks in the standard flavor conserving 2HDMs, and flavorful 2HDMs. By analyzing these four models we obtain a representative overview of the phenomenology of the blind spots in 2HDM+S models, and how they are affected by the choice of flavor structure.

\begin{figure}[tb]
\begin{center} 
\includegraphics[width=0.47\textwidth]{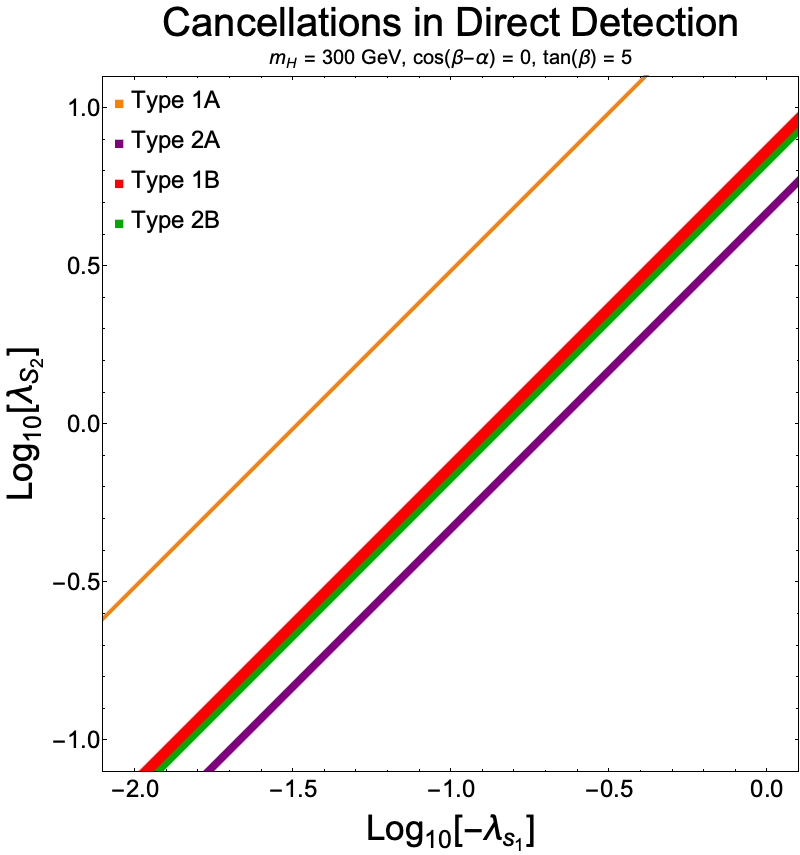}
\caption{Bands corresponding to the position of the direct detection blind spot in the plane of the quartic scalar couplings $\lambda_{S_1}$ and $\lambda_{S_2}$. The finite widths of the shown bands correspond to a variation of the nuclear form factors by $1\sigma$. We show the blind spot regions in four types of 2HDMs: type~1A (orange),~2A (purple),~1B (red) and~2B (green).}
\label{fVal}
\end{center}
\end{figure}

In fig.~\ref{fVal} we show where the direct detection cancellation arises in the $\lambda_{S_1}$ vs $\lambda_{S_2}$ plane for an exemplary choice of the other model parameters:
$m_H = 300$\,GeV, $\cos(\beta-\alpha) = 0$, $\tan\beta = 5$, and $\lambda_{S_{12}} = 0$.
The choice $\cos{(\beta-\alpha)} = 0$ (or more generally,  $\cos{(\beta-\alpha)} \ll 1$) corresponds to SM-like couplings of the light Higgs boson $h$. This is motivated by the good agreement of Higgs couplings measurements at the LHC with SM predictions~\cite{Khachatryan:2016vau,Sirunyan:2018koj,Aaboud:2018urx,Sirunyan:2018hoz,Aaboud:2018zhk,Sirunyan:2018kst,Sirunyan:2017khh,Aaboud:2018pen}. 
Setting the coupling $\lambda_{S_{12}}$ to zero can be enforced by a Peccei-Quinn type symmetry acting on the Higgs doublets~\cite{Peccei:1977ur}.

The width of the bands in fig.~\ref{fVal} correspond to a $1\sigma$ variation of the nucleon form factors~\cite{Belanger:2008sj}. For type 1A models the cancellation occurs for a larger hierarchy between $\lambda_{S_1}$ and $\lambda_{S_2}$ as compared to other types of models. This is because all the couplings to the heavy Higgs are sub-leading in this model, causing the cancellation between the diagrams occurring at smaller values of $g_{SSh}$, and hence generally smaller values of $\lambda_{S_1}$. The other three flavor structures all have similar values for quartic couplings in the cancellation regions as the heavy Higgs plays a larger role forcing $g_{SSh}$ to take on larger values than in the type~1A model.
The precise location of the cancellation regions also depends on the choice of $\tan\beta$ and $m_H$. Larger values of $m_H$ generically require larger values of $\lambda_{S_2}$ for the cancellation to occur.

Note that the cancellation arises if one of the two quartic couplings $\lambda_{S_1}$ or $\lambda_{S_2}$ are negative. Negative terms in the potential can lead to the potential being unbounded from below, meaning there could exist field directions for which the potential goes to negative infinity.
 To study this possibility, we parameterize the three neutral scalar directions as follows:
 \begin{eqnarray}
 S=R\cos\theta\\
 \phi_1^0=R\sin\theta\cos\phi\\
 \phi_2^0=R\sin\theta\sin\phi
 \end{eqnarray}
 and study the positivity of the largest powers of $R$, which is $R^4$, in the potential on the sphere defined by the angles $\theta,\phi$. The requirement that the potential be positive as $R\to\infty$ then reads:
\begin{eqnarray}
    \lambda_{S_1} \sin{\theta}^2 \cos{\theta}^2 \cos{\phi}^2 + \lambda_{S_2}\sin{\theta}^2 \cos{\theta}^2 \sin{\phi}^2  +  \lambda_S \cos{\theta}^4 + 2 \lambda_{S_{12}}\sin{\theta}^2 \cos{\theta}^2 \sin{\phi} \cos{\phi} \nonumber \\+\frac{\lambda_1}{2} \sin{\theta}^4 \cos{\phi}^4 + \frac{\lambda_2}{2} \sin{\theta}^4 \sin{\phi}^4 + \lambda_{345} \sin{\theta}^4 \cos{\phi}^2 \sin{\phi}^2>0 ~,
\label{stability}
\end{eqnarray}
where the $\lambda_i$, $i = 1,\dots,5$ are quartic couplings in the 2HDM potential as defined in~\cite{Branco:2011iw} and $\lambda_{345} = \lambda_{3} + \lambda_{4} + \lambda_{5}$.
In the region of interest to us, $\lambda_{S_1}$ takes smaller values compared to $\lambda_{S_2}$. For this reason we take $\lambda_{S_1}$ to be negative; under this assumption and assuming that $\lambda_S$, $\lambda_1$, $\lambda_2$, $\lambda_{345}$ are ${\cal O}(1)$ and positive, then eq.~(\ref{stability}) can be always satisfied, and thus the potential is stable. 

\subsection{Blind Spots in Invisible Higgs Decays}

The second blind spot occurs for invisible Higgs decays. The decay width of the Higgs to the DM is given by
\begin{equation}
    \Gamma(h \to SS) = \frac{ g_{SSh}^2}{32 \pi m_h}\Bigg(1-4\frac{m_S^2}{m_h^2}\Bigg)^{1/2}
    \label{eq:BRinvisibleHiggs}
\end{equation}
From this expression it is clear that we have a blind spot centered around $g_{SSh} = 0$. Using eq.~(\ref{eq:gSSh}) we see that this cancellation occurs when, 
\begin{equation}
 \frac{\lambda_{S_1}}{\lambda_{S_2}} = \frac{\tan{\alpha}}{\tan{\beta}},
\end{equation}
where we have set $\lambda_{S_{12}} = 0$. For simplicity, we keep this choice for remainder of the analysis, but note that this gives no fundamental difference to the analysis. Blind spots exist for any choice of $\lambda_{S_{12}}$, and are simply shifted in the parameter space when $\lambda_{S_{12}} \neq 0$. 

\begin{figure}
\begin{center}
\begin{tabular}{cc}
\includegraphics[width=150mm]{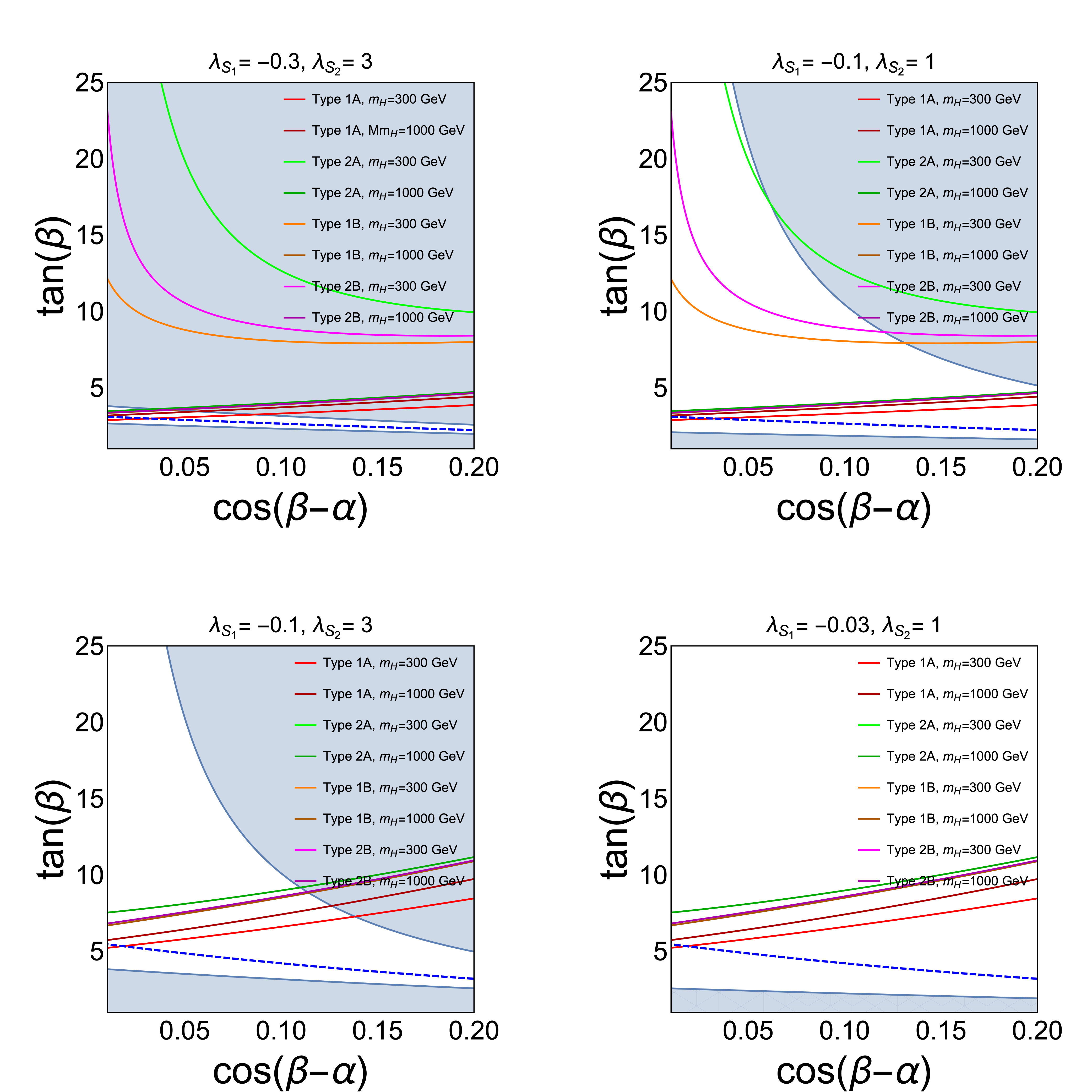} & 
\end{tabular}
\caption{Regions of parameter space in the $\cos{\beta - \alpha}$ vs $\tan{\beta}$ plane with blind spots for invisible Higgs decays for DM mass $m_S = 45$\,GeV and heavy Higgs mass $m_H = 300$\,GeV or $m_H = 1000$\,GeV.
The region excluded by invisible Higgs decays is shaded in blue, with the exact cancellation line in dashed blue. The bound and the cancellation line depend on the choice of $\lambda_{S_1}$ and $\lambda_{S_2}$ but are independent of the 2HDM flavor structure. Overlaid are the direct detection blind spots that occur for our four benchmark 2HDMs.}
\label{overlap}
\end{center}
\end{figure}

 By imposing that the invisible Higgs branching ratio BR$(h \to SS) < 0.19$~\cite{Sirunyan:2018owy}, we find that $g_{SSh}/v$ has to be less than $O(0.1)$ (the exact value changes depending on the choice of $m_S$). We show, in fig.~\ref{overlap}, under which conditions the direct detection cancellations overlap with parameter space where $g_{SSh}$ is sufficiently small to avoid invisible Higgs decay constraints. We show this for the DM matter mass of $m_S =45$\,GeV, with various choices for $\lambda_{S_1}$ and $\lambda_{S_2}$ (corresponding to the four panels in fig.~\ref{overlap}). 
 Invisible Higgs decays exclude the region shaded in blue, with the exact cancellation line in dashed blue. The bound and the cancellation line depend on the choice of $\lambda_{S_1}$ and $\lambda_{S_2}$ but are independent of the type of 2HDM. Overlaid are the direct detection blind spots that occur for our four benchmark 2HDMs for two masses of the heavy Higgs $m_H = 300$\,GeV or $m_H = 1000$\,GeV.
 For a heavy Higgs mass of $300$\,GeV, the direct detection cancellation in the type~2A, type~1B, and type~2B occurs for values of $\tan\beta$ outside the shown plot range.
 
 The type 1A model avoids the constraints most easily as regardless of the parameters of the model the cancellation regions for direct detection and invisible Higgs decay are generally very similar. As mentioned above, in the type~1A model the quarks primarily couple to the SM-Higgs and thus the direct detection cancellation is driven by $g_{SSh}$ being small just like invisible Higgs decay. For the other models we see that generally as $\lambda_{S_1}$ is lowered the bound from the invisible Higgs decays is weakened. However, as we will see below this also generally coincides with regions of parameter space where the DM is overabundant. With this in mind the most promising parameter space for ``double blind spots'' occurs for moderate values of $\lambda_{S_1}$ and $\lambda_{S_2}$.

\subsection{Fine Tuning of the Blind Spots}
\begin{figure}[tb]
\begin{center} 
\includegraphics[width=0.45\textwidth]{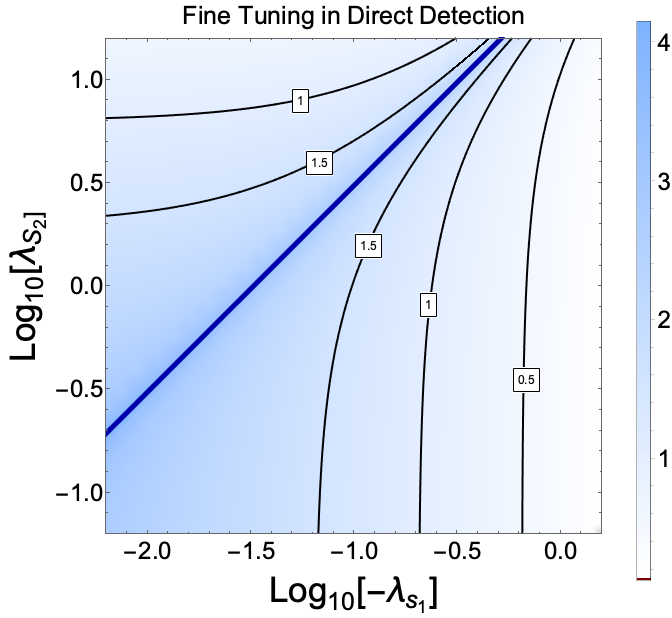}
\includegraphics[width=0.45\textwidth]{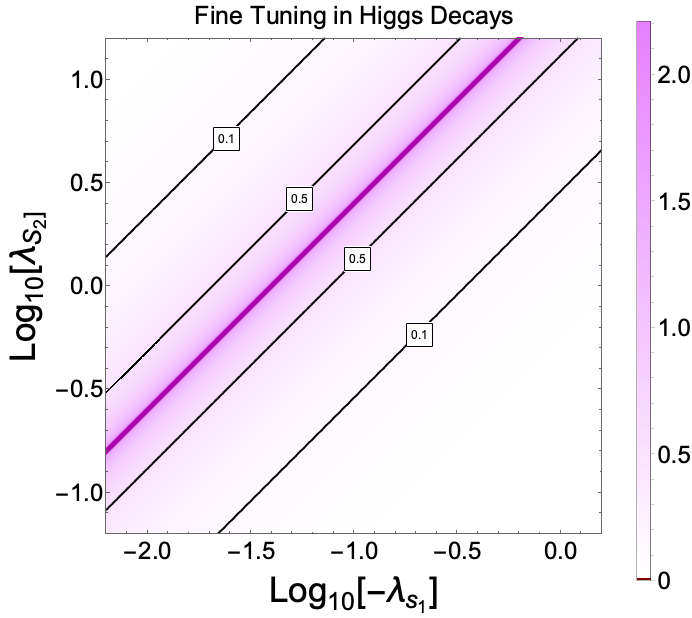}
\caption{The fine tuning of the direct detection blind spot (left) and the invisible Higgs decays blind spot (right) in the $\lambda_{S_1}$ vs. $\lambda_{S_2}$ plane, for the type~1A model. The darker regions represent areas of higher tuning. For direct detection the black lines show the contours for the tuning, and the blue line shows where the exact cancellation lies. For invisible Higgs decays the black lines show the contours for the tuning, and the pink line shows the exact cancellation.}
\label{fineTune}
\end{center}
\end{figure}

The question of how ``natural'' the blind spots we point out are is connected to what extent the parameters must be finely tuned for those blind spots to occur. Fine tuning refers to scenarios in which a single or several parameters must take on very specific values in order for a model to be consistent. The presence of accidental cancellations in our model could be associated with potentially large fine tuning. One way to quantify the fine tuning of a function is to employ the quantity~\cite{Barbieri:1987fn}
\begin{equation}
    g(\vec{x})  = \sum_{i = 1}^n \left| \frac{x_i}{f(x_i)}\frac{\partial f(x_i)}{\partial x_i} \right|~,
\end{equation}
where $g(\vec{x})$ is the amount of tuning in the function $f(\vec{x})$. We show the fine tuning of our model in fig.~\ref{fineTune}, where $x_i=\lambda_{S_i}$, $i=1,2$, considering both the direct detection cross section and the invisible Higgs width. Generally, the tuning is mild in both models, but as expected the tuning gets very large directly at the cancellation lines. As we will discuss later based on current experimental constraints one does not necessarily need to live exactly on this constraint, particularly for higher dark matter masses. So, there is still probable parameter space that does not suffer from large fine tuning. However, for low mass DM the tuning can be quite large. 

\section{Phenomenology of Blind Spots}
\label{pheno} 
\begin{figure}[ht]
\begin{center}
\includegraphics[width=180mm]{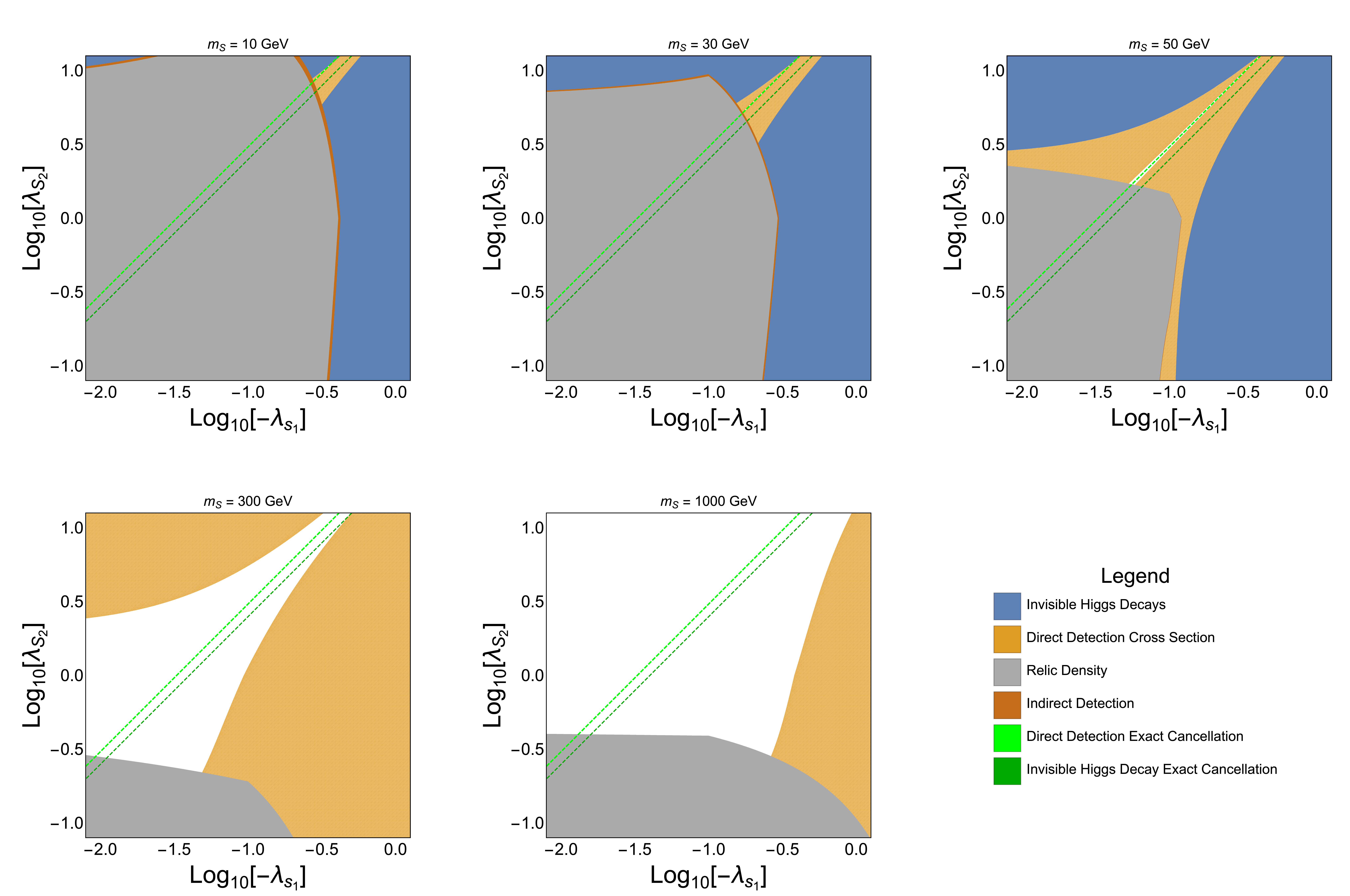}
\caption{Constraints in the $\lambda_{S_1}$ vs $\lambda_{S_2}$ plane in the type~1A model for $\cos(\beta-\alpha) = 0$, $\tan{\beta} = 5$, $m_H = 300$ GeV, and various increasing values of dark matter mass $m_S$. The color coding of the various constraints is specified in the legend.}
\label{varyMass}
\end{center}
\end{figure}
\begin{figure}[ht]
\begin{center}
\includegraphics[width=180mm]{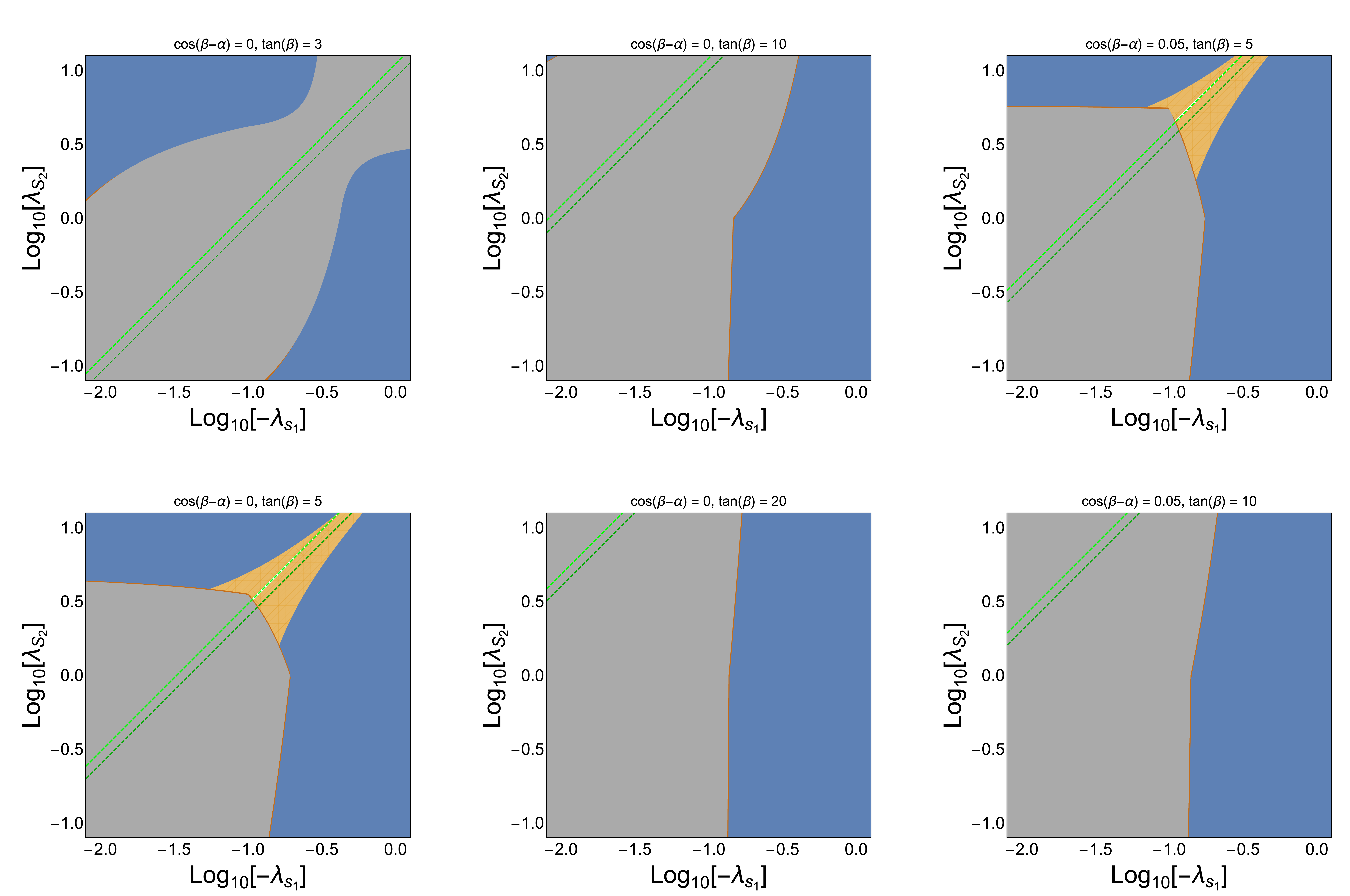}
\caption{Constraints in the $\lambda_{S_1}$ vs $\lambda_{S_2}$ plane in the type~1A model for a dark matter mass $m_S = 45$\,GeV and heavy Higgs mass $m_H = 300$\,GeV, varying the values for $\cos(\beta-\alpha)$, $\tan{\beta}$. The color coding of the constraints is as in fig.~\ref{varyMass}.}
\label{varyAngles}
\end{center}
\end{figure}

In order to better understand the physical parameter space of the blind spots and the resulting phenomenology we study the cancellations in the context of the four 2HDMs discussed above (type~1A, type~2A, type~1B, type~2B). We implement the four models in the micrOMEGAs framework~\cite{Belanger:2013oya}, modifying the default inert doublet model of micrOMEGAs to have the coupling structures under consideration, and use this to calculate the relic density and indirect detection limits. The direct detection cross section and invisible Higgs decays strengths are calculated analytically from the expressions in eq.~(\ref{eq:directdetection}) and eq.~(\ref{eq:BRinvisibleHiggs}).

We explore the parameter space of the quartic couplings $\lambda_{S_1}$ and $\lambda_{S_2}$ for various choices of the dark matter mass $m_S$, the 2HDM parameters $\cos(\beta-\alpha)$ and $\tan\beta$ and the type of 2HDM.

In fig.~\ref{varyMass}, we focus on the type~1A model and vary $m_S = 10, 30, 50, 300, 1000$\,GeV for fixed $\cos(\beta-\alpha) = 0$, $\tan\beta = 5$ (as in fig.~\ref{fVal}).
The white regions are allowed by all constraints.
We see that generally as the DM mass is increased the constraints on the model are weakened. The two phenomenologically distinct regions of parameter space are when the DM mass is below and above half the Higgs mass. If $m_S > \frac{1}{2}m_h$ the constraints from invisible Higgs decays are automatically avoided. 
The constraints from direct detection are also particularly strong for the chosen lighter dark matter masses, $m_S = 10, 30, 50$\,GeV. For those masses only a thin band close to the direct detection blind spot corresponds to viable parameter space.
With this in mind, in the following we consider two benchmark masses of $m_S = 45$ GeV and $m_S = 300$ GeV. 

In fig.~\ref{varyAngles} we show how the parameter space varies for different angles $\cos(\beta-\alpha)$ and $\tan{\beta}$. We observe that moderate values of $\tan{\beta}$ are favorable for these scenarios. As $\tan{\beta}$ gets small the couplings of the quarks to the heavy Higgs increases (for the type 1A model), this causes destructive interference between the annihilation channels of DM through SM-like and heavy Higgs which constrains small values of $\tan{\beta}$ (this is specific to the type 1A model).  We do not find viable parameter space for $\tan\beta \gtrsim 10$ due to stronger constraints from direct detection and the relic density. This gives us a sweet spot for moderate values of $\tan{\beta}$ where the DM can efficiently annihilate in the early universe. $\cos(\beta-\alpha)$ has only a small impact on the results, making the relic density only slightly more constraining. Considering this along with the constraints on the 2HDM parameter space, as shown in ~\cite{Altmannshofer:2018bch}, we focus on the benchmark case of $\cos(\beta-\alpha) = 0, \tan{\beta} = 5$ in the following.

\begin{figure}[ht]
\begin{center}
\begin{tabular}{cc}
\includegraphics[width=65mm]{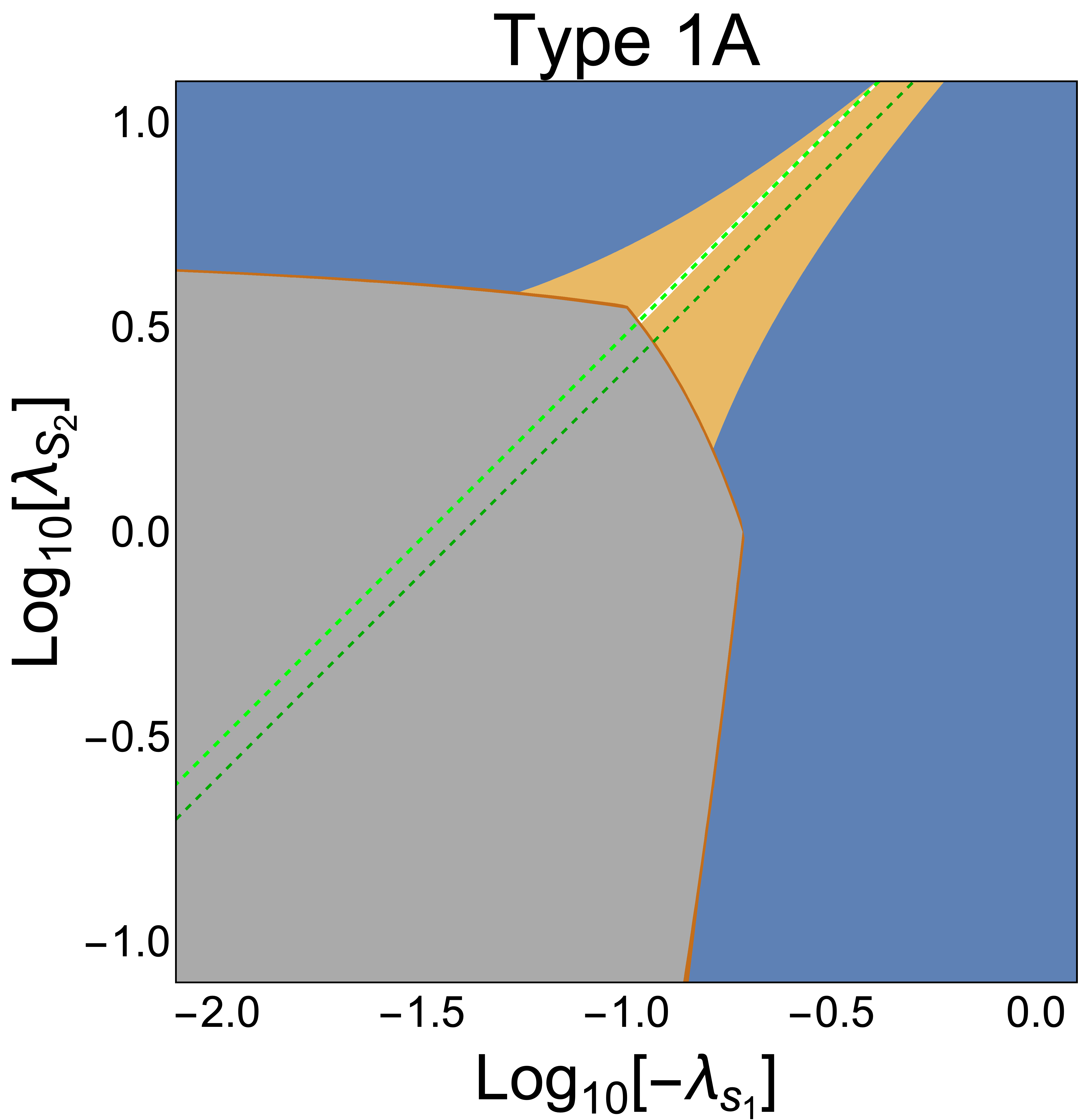} &   \includegraphics[width=65mm]{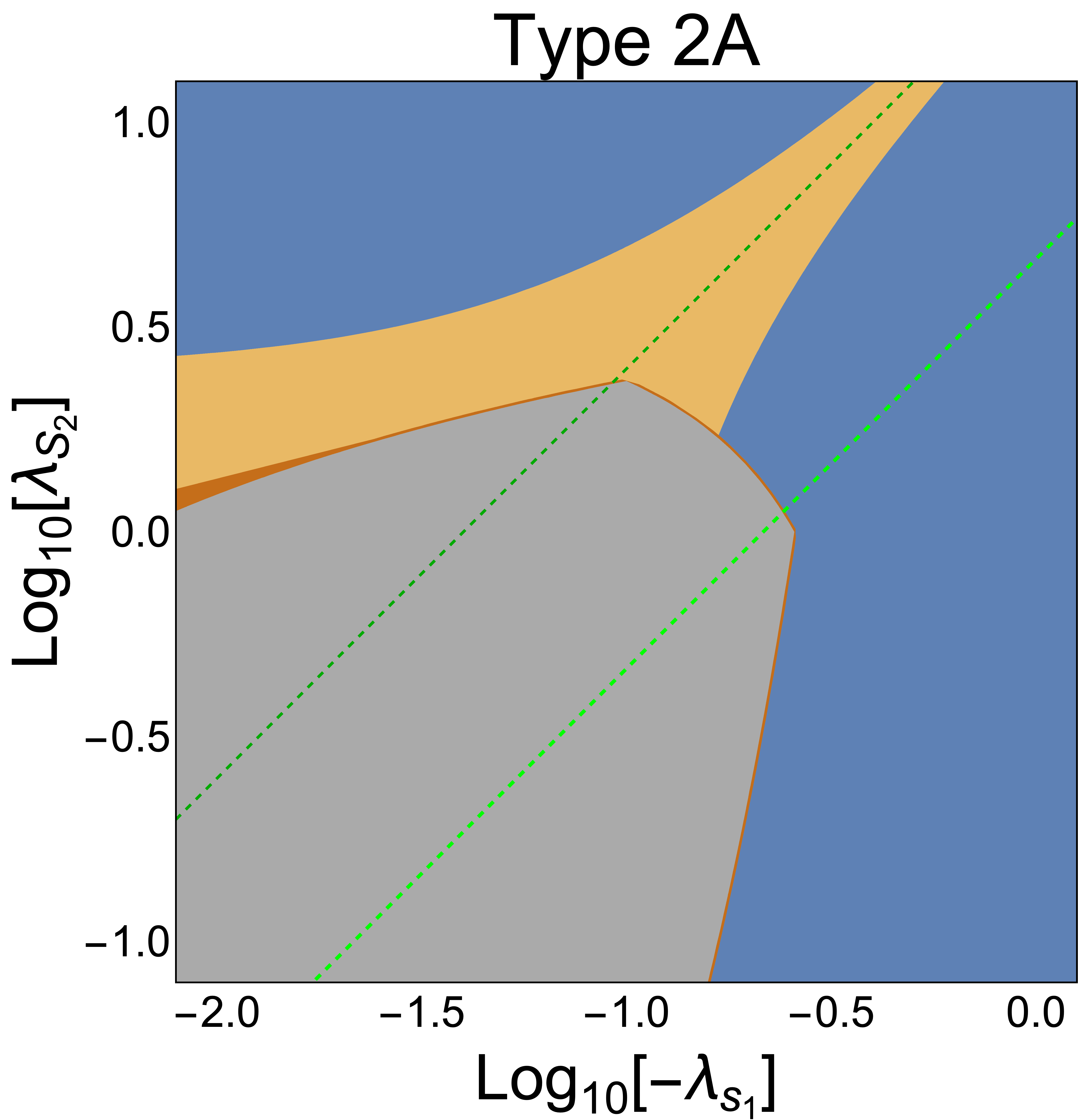}\\
\includegraphics[width=65mm]{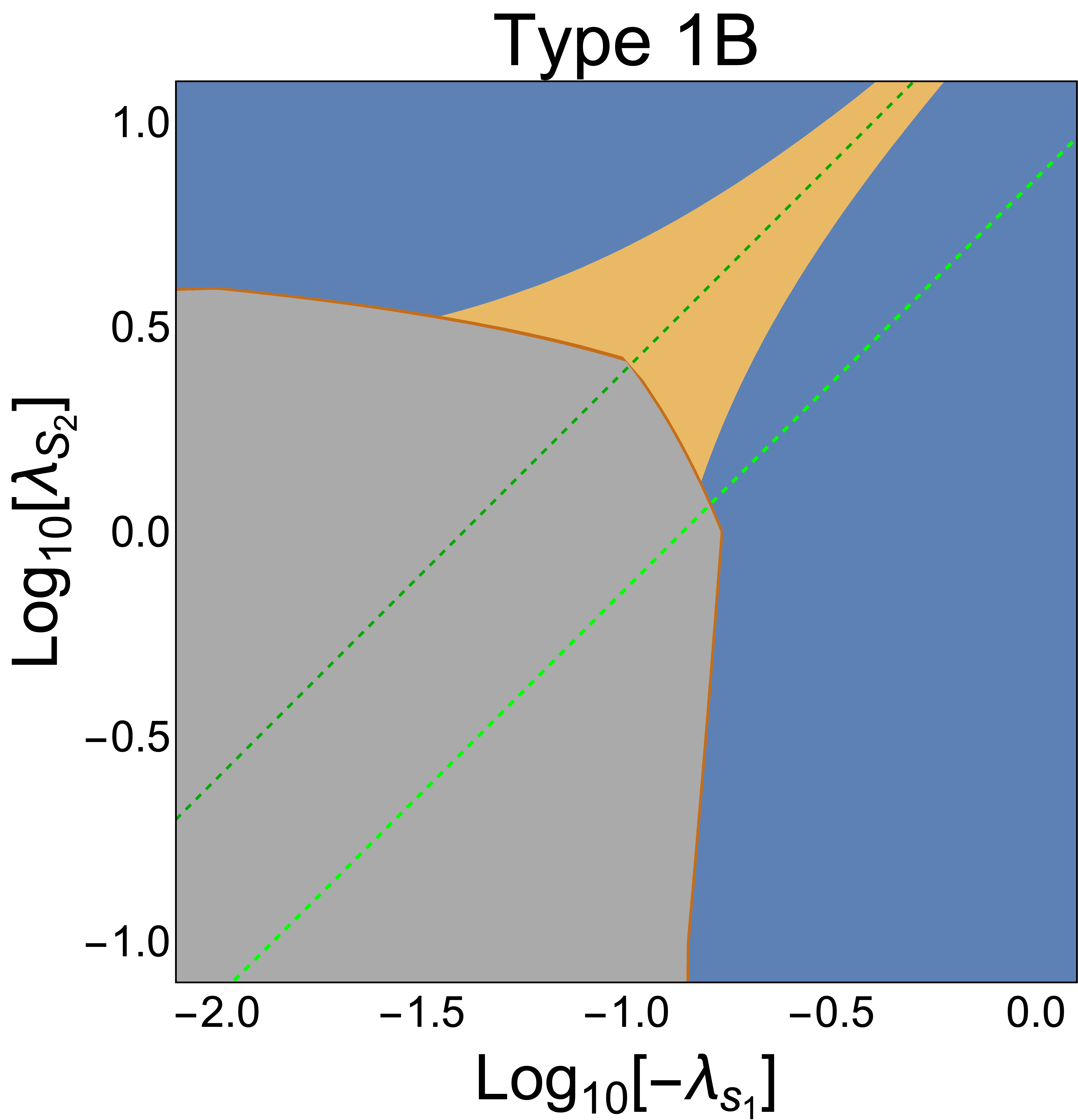} &   \includegraphics[width=65mm]{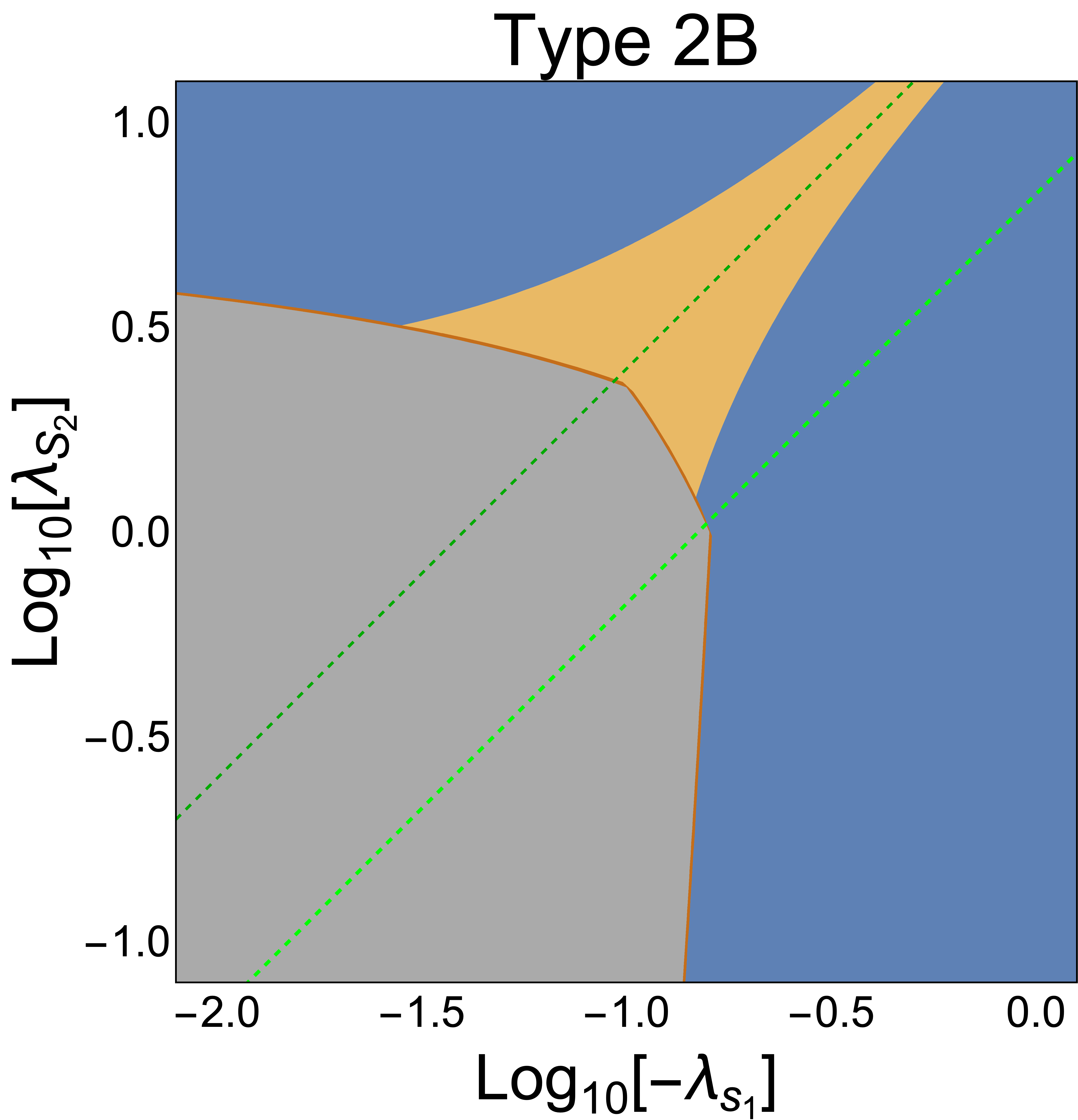}\\
\end{tabular}
\caption{Constraints in the $\lambda_{S_1}$ vs $\lambda_{S_2}$ plane for dark matter mass $m_S = 45$\,GeV, heavy Higgs mass $m_H = 300$\,GeV, $\cos(\beta-\alpha) = 0$, and $\tan{\beta} = 5$. We show type~1A (top left), type~2A (top right), type~1B (bottom left), and type~2B (bottom right). The color coding of the constraints is as in fig.~\ref{varyMass}. } 
\label{mX45cos0}
\end{center}
\end{figure}

In fig.~\ref{mX45cos0} we show the constraints for fixed dark matter mass $m_S = 45$ in the different types of 2HDMs. In the type~2A, type~1B, and type~2B models, the low DM mass regions are still highly constrained by the combination of relic density, direct detection and invisible Higgs decays. In particular, in the region where DM is not overabundant the direct detection blind spots and the invisible Higgs blind spots do not overlap in these types. Only in the type~1A model we have a viable doubly blind spot, where direct detection and invisible Higgs decays are simultaneously avoided. However, in order for this doubly blind spot to occur we see that the fine tuning must be quite high as shown in, fig.~\ref{fineTune}. Although the doubly blind spot only occurs for one of the benchmark models it is still in stark contrast to simple SM+S WIMP models where this region is ruled out by both direct detection and invisible Higgs decays. 

Future experiments prove promising for the remaining parameter space of this low mass dark matter. Projections from the HL-HLC show a future sensitivity of BR$(h \to SS) < 0.025$~\cite{Cepeda:2019klc}, and future direct detection experiments which will improve the measurement of the nucleon cross section by more than an order of magnitude \cite{Liu:2017drf, Kang:2018odb}. Given this sensitivity nearly all masses such that $m_S<\frac{1}{2}m_h$ will be probed. 

\begin{figure}
\begin{center}
\begin{tabular}{cc}
\includegraphics[width=65mm]{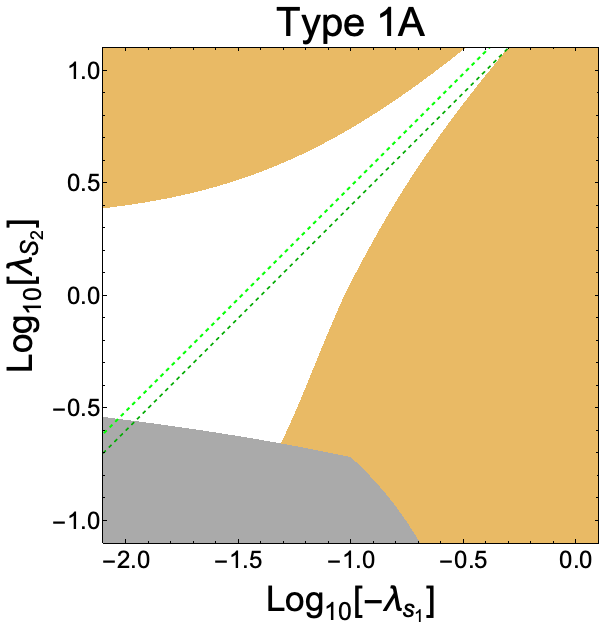} &   \includegraphics[width=65mm]{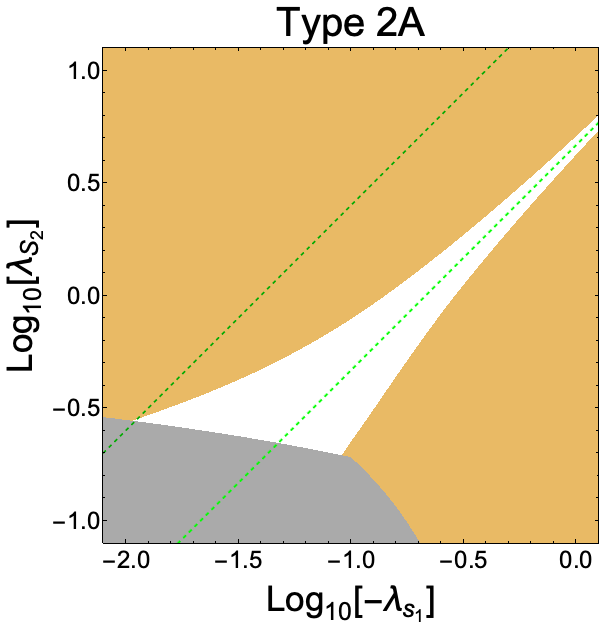}\\
\includegraphics[width=65mm]{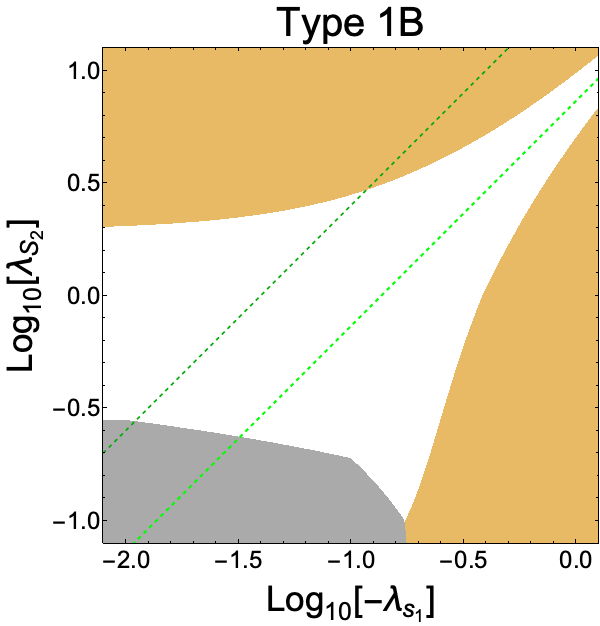} &   \includegraphics[width=65mm]{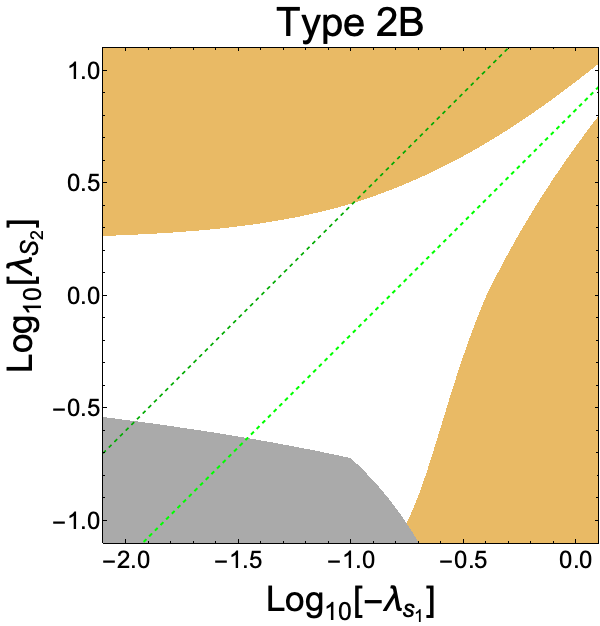}\\
\end{tabular}
\caption{Constraints in the $\lambda_{S_1}$ vs $\lambda_{S_2}$ plane for $m_S = 300$\,GeV, $m_H = 300$GeV, $\cos(\beta-\alpha) = 0$, and $\tan{\beta} = 5$.
We show type~1A (top left), type~2A (top right), type~1B (bottom left), and type~2B (bottom right). The color coding of the constraints is as in fig.~\ref{varyMass}.}
\label{mX300cos0}
\end{center}
\end{figure}

Finally the benchmark case of a heavy dark matter mass $m_S = 300$\,GeV is shown in fig.~\ref{mX300cos0} for the four types of 2HDMs. In this high mass region DM direct detection constraints are alleviated in a much larger portion of parameter space for all four flavor structures and the invisible Higgs constraint is completely absent. This shows that for a variety of flavor structure when the DM mass is high we can expect the direct detection blind spot to open up a large portion parameter space. This blind spot becomes more confined as one lowers the DM mass and more generous as one increases the mass. Similarly, the relic density constraints are much weaker for the high mass DM candidates. Overall, this leads to some viable parameter space in all four benchmark models. Additionally, the viable parameter space in the high DM mass regime can exist quite far from the cancellation lines, where the fine tuning is low, unlike the low mass case where the viable parameter space only occurs in high fine tuning regions. We also see that the non-standard flavor structure of the type~B models allow for an even more generous parameter space than the traditional flavor diagonal structures for the high mass DM benchmark. As discussed in the low mass case, future direct detections will continue to constrain this parameter space, however without the additional constraint from invisible Higgs decays there will still be a large amount of  parameter space. 
 
It is important to remember that for all DM masses explored above, the SM+S model is already excluded by either direct detection or invisible Higgs decays (outside of the resonant region). So, by adding a second Higgs doublet we provide regions of parameter space where DM candidates can exist at much lower masses than are possible in the SM+S case. Additionally, although we explored several fixed flavor structures the cancellations can occur for any generic flavor structure, and are in no way associated only to the structures considered here. 

\section{conclusions}
\label{conclusion}

The WIMP DM paradigm is a simple explanation to the question of the nature of the dark matter in the universe. However, such paradigm has come in recent years under greater and greater pressure as a result of constraints from direct detection experiments as well as from results on the invisible Higgs decay modes. In this work we presented a model where one can take advantage of a second Higgs doublet in order to evade the constraints which invalidate most regions of parameter space of simpler WIMP models based on the existence of a singlet scalar field. 

In particular, we find at DM masses below half the Higgs mass $m_S < \frac{1}{2}m_h$ that one can evade both direct detection and invisible Higgs decay constraints for flavor structures that are type 1A-like as a result of generic {\it blind spots} producing exact or approximate cancellations. Such cancellations depend in detail on the choice of the 2HDM parameters $\cos(\beta - \alpha)$, $\tan{\beta}$ and $m_H$, but generally persist when the couplings of the fermions are primarily associated to the SM-like Higgs. 

We also consider the scenario where the DM mass is large, $m_S>\frac{1}{2}m_h$, where we find that direct detection can be avoided for all considered 2HDM flavor structures. This primarily arises because one no longer needs to avoid the constraints imposed by invisible Higgs decays. Generically, we see that as the dark matter mass increases, the parameter space further opens up. There is also a weak dependence on the choice of $\cos(\beta - \alpha)$, $\tan{\beta}$ and $m_H$. Mostly these choices affect the constraint of the relic density. Smaller values of $\cos(\beta - \alpha)$ and $\tan{\beta}$ typically result in a more open parameter space. 

Overall, we find that with the inclusion of a second Higgs doublet one can access a much larger range of DM masses than in simpler models. This depends somewhat on the flavor structure of these models; however, in all the flavor structures considered, blind spots that facilitate the evasion of direct detection and collider constraints do exist, and, more generally, as we showed, blind spots can exist in any generic 2HDM setup.

\section*{Acknowledgements}

The research of WA and BM is supported by the National Science Foundation under Grant No. PHY-1912719. SP is partly supported by the U.S.\ Department of Energy grant number de-sc0010107.

\bibliographystyle{ieeetr}
\bibliography{refs}

\end{document}